\documentclass[12pt]{iopart}
\usepackage{graphicx}
\usepackage{color}
\usepackage{verbatim}

\begin{document}

\title{On the distillation and purification of phase-diffused squeezed states}

\author{B Hage$^1$, A Franzen$^1$, J DiGuglielmo$^1$, P Marek$^2$, J Fiur\'{a}\v{s}ek$^3$ and R~Schnabel$^1$}

\address{$^1$ Max-Planck-Institut f\"{u}r Gravitationsphysik (Albert-Einstein-Institut) and \\
Leibniz Universit\"{a}t Hannover, Callinstr. 38, 30167 Hannover, Germany}
\address{$^2$ School of Mathematics and Physics, The Queen's University, Belfast BT7
1NN, United Kingdom}
\address{$^3$ Department of Optics, Palack\'{y} University, 17. listopadu 50, 77200 Olomouc, Czech Republic}
\ead{boris.hage@aei.mpg.de}

\begin{abstract}
Recently it was discovered that non-Gaussian decoherence processes,
such as phase-diffusion, can be counteracted by purification and
distillation protocols that are solely built on Gaussian operations.
Here we make use of this experimentally highly accessible regime,
and provide a detailed experimental and theoretical analysis of
several strategies for purification/distillation protocols on
phase-diffused squeezed states. Our results provide valuable
information for the optimization of such protocols with respect to
the choice of the trigger quadrature, the trigger threshold value
and the probability of generating a distilled state.
\end{abstract}
\submitto{\NJP}

%         %%%%%%%%%%%%%%%%%%%%%%%%%%%%%%%%%%%%%%%%%%%%%%
%         %%                                                         INTRODUCTION                                                                   %%
%         %%%%%%%%%%%%%%%%%%%%%%%%%%%%%%%%%%%%%%%%%%%%%%

\section{Introduction}

Continuous variable quantum information processing \cite{Braunstein05} represents
an appealing alternative to the traditional qubit based approaches.
Continuous quantum variables encoded in quadrature components of light
modes \cite{Ou92,SBTBRL03,Grosshans03,Furusawa98,Bowen03a} or collective atomic
spin \cite{Julsgaard01,Julsgaard04,Sherson06} offer several distinct advantages.
Many important quantum information processing primitives could be implemented
deterministically with only linear optics, optical parametric amplifiers (``squeezers'')
and balanced homodyne detection. This includes for example preparation of entangled
two-mode squeezed states \cite{Ou92,Silberhorn01,Schori02,Bowen02,BSLR03}, quantum
teleportation \cite{Furusawa98,Bowen03a,BTBSSRL03}, dense coding \cite{Peng03},
entanglement swapping \cite{Takei05} and quantum cloning \cite{Andersen05}.
Quantum states of light beams could be deterministically stored in an atomic
quantum memory \cite{Julsgaard04} and teleportation of quantum state of a light mode
onto the atomic ensemble has been demonstrated \cite{Sherson06}.

In all these developments, Gaussian states and Gaussian operations
play a prominent role. Recall that Gaussian states are states with
Gaussian Wigner function and Gaussian operations are those quantum
transformations which preserve the Gaussian shape of the Wigner
function. Passive linear optics, squeezers and homodyne detection
are all examples of Gaussian operations and they allow to prepare
arbitrary Gaussian states from coherent states. However, it has been
realized that Gaussian operations are insufficient for entanglement
distillation of Gaussian states \cite{Giedke02,Eisert02,Fiurasek02}.
This is a significant limitation, because entanglement distillation
\cite{Bennett96,Deutsch96} is critical for suppression of noise and
losses which inevitably arise during the distribution of
entanglement between distant parties.  The entanglement
purification/distillation protocol extracts a highly entangled pure state
from many copies of mixed weakly entangled states using only local operations
and classical communication between the parties sharing the states.
It is anticipated that any long-distance quantum
communication network will have to involve entanglement distillation in some form.
A similar no-go theorem has been  proved
also for the concentration of squeezing \cite{Kraus03}. Namely,
given an arbitrary $N$-mode Gaussian state, it is not possible to
increase its squeezing strength by passive Gaussian operations.
Here, the squeezing strength is quantified by the lowest eigenvalue
of the covariance matrix of the state, and passive Gaussian
operations are those building on linear optics, homodyning and
feedforward. This implies that it is impossible to extract in this
way from $N$ copies of a single-mode squeezed Gaussian state a state
with higher squeezing.

Continuous-variable entanglement distillation thus requires to go
beyond the realm of Gaussian states and operations in at least a
single step \cite{Duan00,Opatrny00,Browne03}. A complete protocol
for entanglement distillation of Gaussian states has been suggested
by Browne, Eisert \emph{et al.} \cite{Browne03,Eisert04}. This
protocol consists of two steps. First, the state is de-Gaussified
e.g. by conditional subtraction of a single photon
\cite{Opatrny00,Ourjoumtsev06,Nielsen06,Ourjoumtsev06b,Wakui06}.
Second, the de-Gaussified states are Gaussified using interference
on beam splitters, Gaussian measurements and conditioning. If,
however, the dominant noise itself is such that it de-Gaussifies the
state, then no further de-Gaussification is necessary and the
squeezed or entangled states can be distilled and purified solely by
means of passive linear optics, homodyne detection and conditioning
\cite{Fiurasek05,Franzen06,Heersing06}. Purification of squeezed
states has been demonstrated in this regime in \cite{GAFBL06}.
However, the purification was accompanied with a degraded degree of
squeezing. Recently, we have proposed and experimentally
demonstrated a protocol for purification \emph{and} distillation of
squeezed states suffering from phase fluctuations \cite{Franzen06}.
These fluctuations, which are an important source of noise in
optical communication links, can be suppressed by interference of
two copies of the non-Gaussian phase-diffused squeezed state on a
balanced beam splitter, followed by homodyne detection on one of the
outputs and acceptance or rejection of the resulting state depending
on the measurement outcome. Our procedure works for both single-mode
and two-mode squeezed states and the latter could be distilled by
means of local operations and measurements and classical
communication. The purification/distillation procedure could be
applied iteratively and could serve as a gaussification part of the
universal CV entanglement distillation protocol based on a sequence
of de-Gaussification and Gaussification steps.

In this paper we report on a detailed theoretical and experimental
analysis of several strategies for purification/distillation of
phase-diffused single-mode squeezed states. Our results provide a
detailed insight into the features and properties of the
purification/distillation protocol and represent a significant step
towards the experimental demonstration of continuous-variable
entanglement distillation. Compared to our previous works
\cite{Fiurasek05,Franzen06}, we do not assume only conditioning on
measurements of the initially squeezed amplitude quadrature $x_1$
but allow for detection of arbitrary quadrature component. We find
that the purification/distillation protocol exhibits a far greater
richness than expected. Surprisingly enough if, after the
interference of the two noisy copies of the state on a beam
splitter, we measure the originally anti-squeezed phase quadrature
$p_1$ of the first output beam and condition on $|p_1|<Q$  then this
allows an enhancement of the squeezing of the amplitude quadrature
$x_2$ of the second output beam. This is somewhat unexpected since
one could naively argue that conditioning on small $p_1$ should
enhance the fluctuations of $x_2$, see Sec. \ref{sectwothree}. We
term this technique conjugate purification (CP). In certain regimes
this method is even optimal and  for a given amount of distilled
squeezing yields higher purification/distillation rates than
squeezed quadrature-conditioning. Conditioning on the anti-squeezed
quadrature most clearly reveals the quantum nature of our
purification/distillation protocol.

The paper is organized as follows. In Section \ref{conjugatesection}
we present our purification/distillation protocol for phase-diffused
squeezed states. This section includes its experimental
demonstration as well as its theoretical description, and also a
comparison of experimental results and theoretical predictions. In
Section \ref{sectionthree} we investigate quantum channel probing as
a classical enhancement of our purification/distillation protocol.
Finally, Section \ref{sectionfour} contains a brief summary and
conclusions.

%\begin{figure}
%\centerline{\psfig{figure=squeezingpurefig1.eps,width=0.95\linewidth}}
%\caption{Setup for purification of single-mode squeezed states.}
%\label{schemefig}
%\end{figure}

%         %%%%%%%%%%%%%%%%%%%%%%%%%%%%%%%%%%%%%%%%%%%%%%
%         %%                                   THE PURIFICATION PROTOCOL                                                           %%
%         %%%%%%%%%%%%%%%%%%%%%%%%%%%%%%%%%%%%%%%%%%%%%%

\section{The purification/distillation protocol}
\label{conjugatesection}

\begin{figure}[t]
\begin{center}
\includegraphics[width=100 mm,angle=-90]{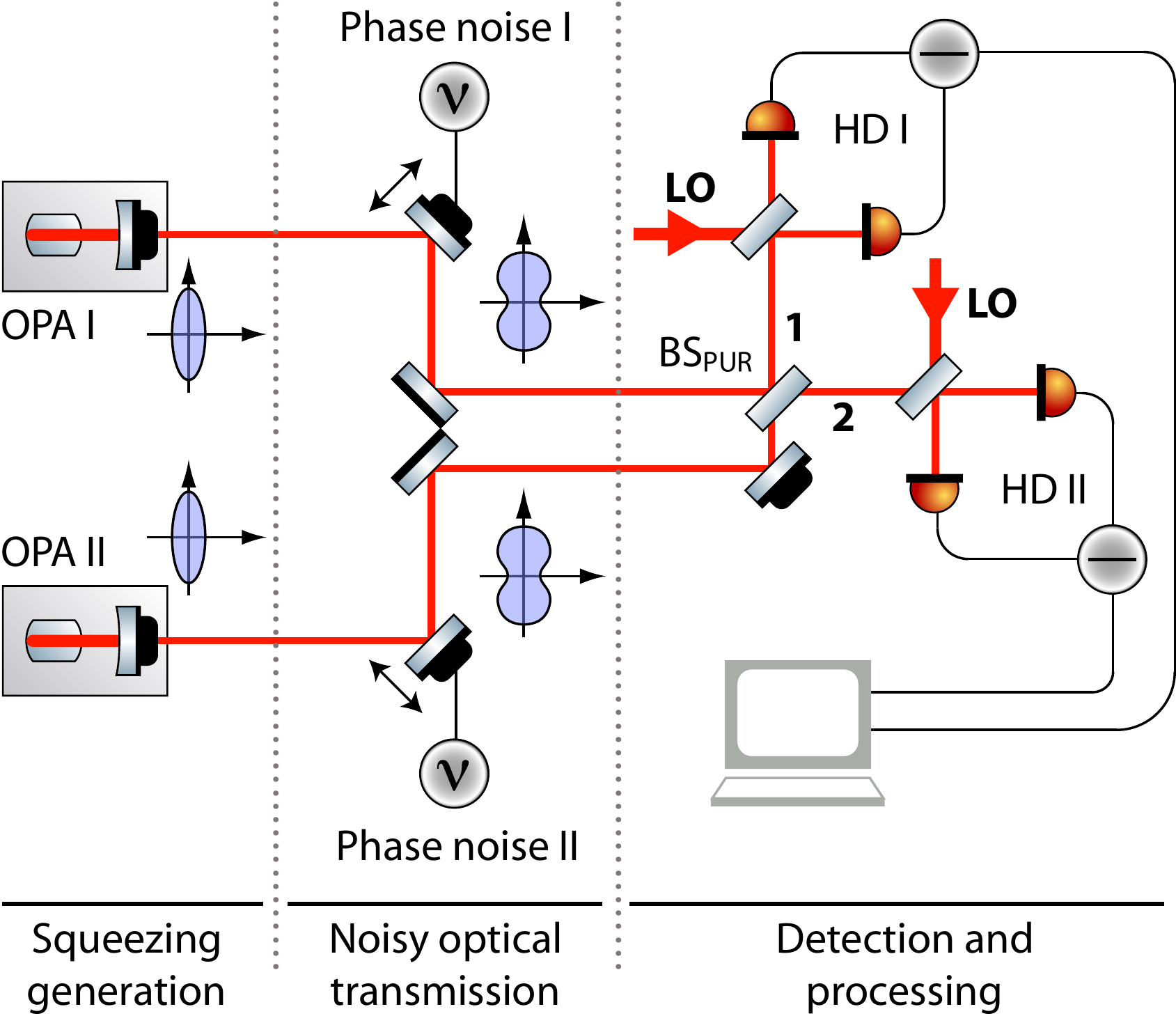}
\end{center}
\caption{ Schematic experimental setup for the demonstration of
purification/distillation of phase-diffused squeezed states. Two
optical parametric amplifiers (OPAI and OPAII) produced one
amplitude-squeezed beam each. Two piezo-electric transducers drove
mirrors to induce random,  gaussian-weighted phase shifts to each
beam to mimic the effect of a noisy optical transmission. The two
phase-diffused beams were then overlapped on a balanced (50/50)
beamsplitter BS$_{\mathrm{PUR}}$. Two homodyne detectors HD\,I and
HD\,II in combination with a digital data acquisition system
synchronously recorded time series of measured quadrature values.
HD\,II serves the purpose of verification and can be replaced by an
arbitrary experiment, which requires a squeezed input beam.}
\label{schemefig}
\end{figure}

The scheme for purification and distillation of phase-diffused
single-mode squeezed states is shown in Fig.\,\ref{schemefig}. It is
a variant of the scheme originally proposed by Browne \emph{et al.}
for Gaussification of continuous-variable entangled states
\cite{Browne03}. Two copies of a single-mode squeezed state
generated in  optical parametric amplifiers (OPA) I and II
independently pass through a noisy quantum channel which imposes
independent and random phase fluctuations upon each state. These
fluctuations reduce the quadrature squeezing and the purity of the
state and make it non-Gaussian \cite{Franzen06}. For the sake of
specificity, we assume that the phase fluctuations exhibit Gaussian
distribution with zero mean and variance $\sigma^2$,
\begin{equation}
\Phi(\phi)= \frac{1}{\sqrt{2\pi \sigma^2}}
\exp\left({-\frac{\phi^2}{2\sigma^2}}\right).
\label{Phiphi}
\end{equation}
Note, however, that the exact form of the phase distribution
$\Phi(\phi)$ is not essential and our findings presented below
qualitatively hold independently of the exact form of $\Phi(\phi)$.
After random phase shifts $\phi_1$ and $\phi_2$ the two squeezed
beams interfere on a balanced beam splitter BS$_{\mathrm{PUR}}$
which forms the core of our purification/distillation setup. After
interference on BS$_{\mathrm{PUR}}$, one output beam impinges on a
balanced homodyne detector HD\,I, which measures a certain
quadrature operator $q_1(\theta)=x_1\cos\theta +p_1\sin\theta $ of
the first output mode. The purification/distillation succeeds and
the state in the output mode $2$ is accepted if $|q_1|<Q$. Otherwise
the purification/distillation fails and the state in mode $2$ is
rejected. In the experiment, the purified state in mode 2 is
monitored using homodyne detector HD\,II which allows us to
determine the performance of the purification/distillation and
verify its success. Our probabilistic purification and distillation
scheme reduces the phase noise, enhances the squeezing and purity of
the state and gaussifies it.  The distinct advantage of our approach
is that it relies solely on passive linear optics and balanced
homodyne detection which is highly efficient and much easier to
implement compared to other approaches such as projections onto
vacuum state by conditioning on no-clicks of single-photon
detectors.

\subsection{Experimental setup}
\label{sectwoone}

Let us now describe our experimental setup in more detail. The
sources of our squeezed states were two optical parametric
amplifiers (OPAs) operated below threshold. Each device was
constructed from a type I noncritically phase-matched MgO:LiNbO$_3$
crystal. The crystals were situated inside hemilithic standing-wave
resonators formed by a high-reflection coated crystal surface on one
end (reflectivity at 1064\,nm $r^2 > 0.999 $) and an externally
mounted outcoupling mirror (reflectivity at 1064\,nm $r^2 = 0.947$) on
the other end. The intracavity surfaces of the nonlinear crystals
were antireflection coated for both the fundamental wavelength
(1064\,nm) and the second harmonic pump field (532\,nm) of the
parametric process. Our laser source was a 2\,W continuous wave
non-planar Nd:YAG ring laser. About 1.7\,W of this laser beam was
frequency doubled to $532 \textrm{ nm}$. Parts of this field were
injected into the OPAs through the outcoupling mirrors for pumping
the nonlinear process. Both OPA cavities were kept on resonance for
the $1064 \textrm{ nm}$ laser radiation. For this purpose
radio-frequency phase modulated control fields on the fundamental
wavelength were injected into the OPA cavities through the
high-reflection coated crystal surfaces.  The reflected control
fields were detected and provided feedback control error signals for
both OPA cavity lengths as well as for both pump field phases. The
amplified error signals were then fed back to piezo-electric
transducers (PZTs) to precisely position the OPA cavity outcoupling
mirrors and mirrors in the optical paths of the pump fields,
respectively.
This enabled the stably controlled production of two beams with
broadband amplitude quadrature squeezing. A nonclassical noise power
reduction of more than $5 \textrm{ dB}$ was directly observed using
homodyne detection. The squeezing spectrum was almost white over a
wide range of frequencies between approximately $5 \textrm{ MHz}$
and $15 \textrm{ MHz}$. Outside this band the performance of the
squeezed light sources was reduced due to phase modulation signals,
technical noise at lower frequencies and the limited OPA cavity
linewidths at higher frequencies.

\begin{figure}[!t!]
\begin{center}
\includegraphics[width = 120 mm]{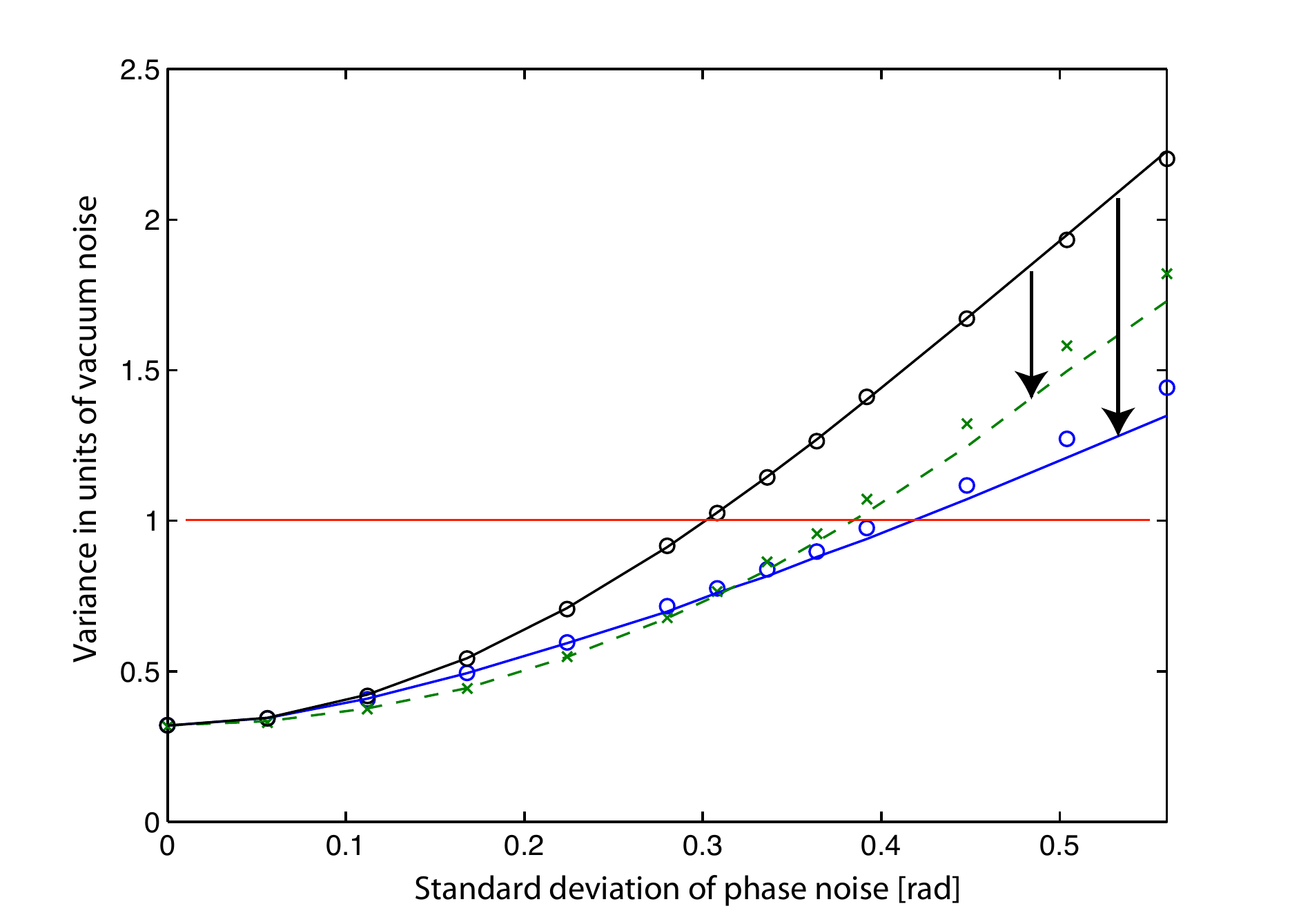}
%\centerline{\psfig{figure=fig2.eps,width=0.90\linewidth}}
\end{center}
\caption{
Demonstration of successful distillation of squeezed states for two different trigger strategies
versus strength of phase fluctuations $\sigma$. Shown are measured variances of the (initially)
squeezed quadrature for the phase diffused input states $V_{\mathrm{in}}$ (top, black)
and for distilled states $V_{\mathrm{out}}$, when conditioned on the squeezed quadrature
$|x_1|<Q$ (circles, blue) and when conditioned on the anti-squeezed quadrature $|p_1|<Q$ (crosses, green),
respectively. Theoretical values are represented by solid and dashed lines.
Here, the trigger threshold was set to $Q=1.0$. For $\sigma < 0.3$ conditioning
on the anti-squeezed quadrature was more efficient than conditioning on the squeezed quadrature.
Red line indicates the shot-noise level.
Without phase diffusion the squeezed state variances were measured to $V_{x} = 0.32$ and $V_{p} = 8.5$.
}
\label{conjugatesigma}
\end{figure}

To realistically mimic the effects of noisy optical transmission
channels each squeezed beam was reflected off a highreflection
mirror that was randomly shifted by a PZT. Hence, random phase
shifts were introduced on the beams as they would occur when
transmitting the beams through optical fibers of considerable
lengths. The voltages driving the PZTs were carefully produced to
meet certain criteria. Although no special form of the noise is in
principle required for a purification/distillation experiment we
wanted to operate in a regime where the noise has well defined
properties and could be easily modeled theoretically. Two
independent random number generators produced data strings with a
white Gaussian distribution. Both strings were then digitally
filtered to limit the noise frequency band to $1$--$5 \textrm{
kHz}$. Performing homodyne detection on each of the beams confirmed
that the amount of squeezing degraded in the same way when the
strength of the phase noise was increased, see Fig.
\ref{conjugatesigma} (black curve/circles). Each homodyne detector
was constructed from a matched pair of ETX-500 high-efficiency photo
diodes. Homodyne visibilities were at $98.4 \%$ for HD\,I and $98.7
\%$ for HD\,II.

In order to demonstrate purification/distillation the two
phase-diffused squeezed beams were overlapped on a balanced (50/50)
beamsplitter. Here, a visibility of 98.2\% was achieved. The
relative phase of the two beams on the beamsplitter was actively
controlled (with a control loop bandwidth \emph{below} the phase
noise frequency band) using phase modulation sidebands present on
the squeezed beams. The output beams from the beamsplitter were then
detected using the homodyne detectors HD\,I and HD\,II, each of
which was servo-controlled to detect the appropriate quadratures
(again with feedback control loop bandwidths \emph{smaller} than the
phase noise frequencies). Each detector difference current was
subsequently electronically mixed with a $7 \textrm{ MHz}$ local
oscillator. The demodulated signals were then filtered with steep
low-pass filters (anti-aliasing filters) at $40 \textrm{ kHz}$ and
synchronously sampled with $100 \textrm{ kHz}$. The nonlinear phase
response of these filters was digitally compensated to regain a
constant group delay. Both time series of quadrature values from
HD\,I and HD\,II, respectively, were then postprocessed to perform
different purification/distillation protocols.

\subsection{Theoretical description}
\label{sectwotwo}

In this section we derive the formula for the squeezed-quadrature
variance $V_{\mathrm{out}}\equiv \langle (\Delta x_2)^2\rangle$  of
the purified output state of our protocol. We consider the general
scenario in which a quadrature
$q_1(\theta)=x_1\cos\theta+p_1\sin\theta$ is measured by HD\,I and a
positive purification/distillation trigger signal is provided if
$|q_1|<Q$. Again, we use $x_j$ and $p_j$ to denote the amplitude or
phase quadrature of the $j$th beam, respectively. We assume that
before phase randomization the $x$ quadrature of each beam is
squeezed and that there are no correlations between $x$ and $p$
quadratures. The covariance matrix of each squeezed mode thus
attains a diagonal form, $\gamma=\mathrm{diag}(V_x,V_p)$, where
$V_x$ and $V_p$ denote variances of $x$ and $p$ quadratures. For the
vacuum state we have $V_x=V_p=1$ and the mode is  squeezed if
$V_x<1$. Recall that $V_x V_p \geq 1$ as a consequence of the
Heisenberg uncertainty relation.

Assume for  a moment that the random phase shifts $\phi_j$ on each mode are fixed. Let
$V_{xj}=V_x\cos^2\phi_j+V_p\sin^2\phi_j$ and $V_{pj}=V_p\cos^2\phi_j+V_x\sin^2\phi_j$,
$j=1,2$,
denote the variances of $x$ and $p$ quadratures of phase-shifted squeezed states
impinging on BS$_{\mathrm{PUR}}$. For a fixed $\phi_j$, the two-mode state
at the output of BS$_{\mathrm{PUR}}$ and prior to conditioning measurement is Gaussian.
Consequently, the joint probability distribution of quadratures  $q_1(\theta)$ of mode 1 and
$x_2$ of mode 2 at the output of BS$_{\mathrm{PUR}}$ is also Gaussian
and  reads,
\begin{equation}
P_{12}(q_1,x_2)= \frac{1}{2\pi
\sqrt{D}}\exp\left[-\frac{Bq_1^2+Ax_2^2-2Cq_1 x_2}{2D}\right].
\end{equation}
Here $D=AB-C^2$, $A$ and $B$ are variances of quadratures $q_1$ and $x_2$ evaluated for the state
at the output of BS$_{\mathrm{PUR}}$,
\begin{eqnarray}
A&=&\frac{V_{x1}+V_{x2}}{2}\cos^2\theta+\frac{V_{p1}+V_{p2}}{2}\sin^2\theta
\nonumber \\
 & & +\frac{V_p-V_x}{4}[\sin(2\phi_1)+\sin(2\phi_2)]\sin(2\theta),
\nonumber \\
B&=&\frac{V_{x1}+V_{x2}}{2}.
\end{eqnarray}
The correlation between the quadratures $C=\langle \Delta q_1 \Delta x_2\rangle$
reads
\begin{equation}
C=\frac{V_{x1}-V_{x2}}{2}\cos\theta
+\frac{V_p-V_x}{4}[\sin(2\phi_1)-\sin(2\phi_2)]\sin\theta.
\end{equation}

The non-normalized distribution of $x_2$ conditional on $|q_1|<Q$ reads
\begin{equation}
P_{\mathrm{cond}}(x_2) = \int_{-Q}^{Q} P_{12}(q_1,x_2) d q_1.
\label{Pcond}
\end{equation}
If the phase fluctuations are symmetric, $\Phi(-\phi)=\Phi(\phi)$ then the mean
value of the quadrature $x_2$ of the purified state is zero and we shall assume that this holds
throughout the rest of the paper. The variance $V_{\mathrm{out}}$  of quadrature $x_2$ then becomes equal to
$\langle x_2^2\rangle$ calculated from the conditional probability distribution
(\ref{Pcond}), averaged over all random phase shifts and properly normalized,
\begin{equation}
V_{\mathrm{out}}= \frac{1}{\mathcal{P}}\int_{\phi_1}\int_{\phi_2}
\int_{-\infty}^\infty x_2^2 P_{\mathrm{cond}}(x_2) d x_2 \Phi(\phi_1)
\Phi(\phi_2) d \phi_1 d \phi_2,
\end{equation}
where
\begin{equation}
{\mathcal{P}}= \int_{\phi_1}\int_{\phi_2} \int_{-\infty}^\infty P_{\mathrm{cond}}(x_2) d x_2 \Phi(\phi_1)
\Phi(\phi_2) d \phi_1 d \phi_2
\end{equation}
is the probability of successful purification/distillation. The integration over $x_2$ can be
explicitly carried out and after some algebra we arrive at
\begin{eqnarray}
V_{\mathrm{out}}&=&\frac{1}{\mathcal{P}}\int_{\phi_1}\int_{\phi_2}
\left[B \, \mathrm{erf}\left(\frac{Q}{\sqrt{2A}}\right)-\sqrt{\frac{2}{\pi}}\frac{C^2
Q}{A^{3/2}} e^{-\frac{Q^2}{2A}} \right]
\Phi(\phi_1) \Phi(\phi_2) d \phi_1 d \phi_2.
\label{Voutconjugate}
\end{eqnarray}
We also obtain a simplified formula for the success probability,
\begin{equation}
\mathcal{P}=
\int_{\phi_1}\int_{\phi_2}
\mathrm{erf}\left(\frac{Q}{\sqrt{2A}}\right)
\Phi(\phi_1) \Phi(\phi_2) d \phi_1 d \phi_2.
\label{Pconjugate}
\end{equation}
For comparison we also calculate the variance of the $x$ quadrature
of the phase-diffused state before purification/distillation,
\begin{equation}
V_{\mathrm{in}}=\int_{\phi}(V_x\cos^2\phi+V_p\sin^2\phi) \Phi(\phi) d\phi.
\end{equation}
Successful purification/distillation increases the squeezing which
is indicated by $V_{\mathrm{out}}<V_{\mathrm{in}}$. Note that
$V_{\mathrm{out}} \geq V_x$ would always hold and the
purification/distillation cannot reduce $V_{\mathrm{out}}$ below the
variance $V_x$ of the original state before transmission through a
noisy channel. However, the method can restore the squeezing lost
due to the phase fluctuations (or other non-Gaussian noise). After
purification/distillation, the state also becomes gaussified and its
purity increases \cite{Fiurasek05,Franzen06}, which clearly
demonstrates that our method meets all requirements imposed on a
proper purification/distillation protocol.

Non-unit detection efficiency $\eta<1$ of the homodyne detectors can
be easily incorporated in our theoretical model by making
substitutions $V_{xj} \mapsto \eta V_{xj}+1-\eta$ and $V_{pj}
\mapsto \eta V_{pj}+1-\eta$. Here we assume for the sake of
simplicity that the detection efficiency is the same for the
triggering detector HD\,I and the verification detector HD\,II which
monitors the purified state. If $V_x$ and $V_p$ are determined by
homodyne detection then these measured values already include the
effects of the non-unit detection efficiency and can be directly
inserted into the above formulas.

\begin{figure}[!t!]
\begin{center}
\includegraphics[width = 120 mm]{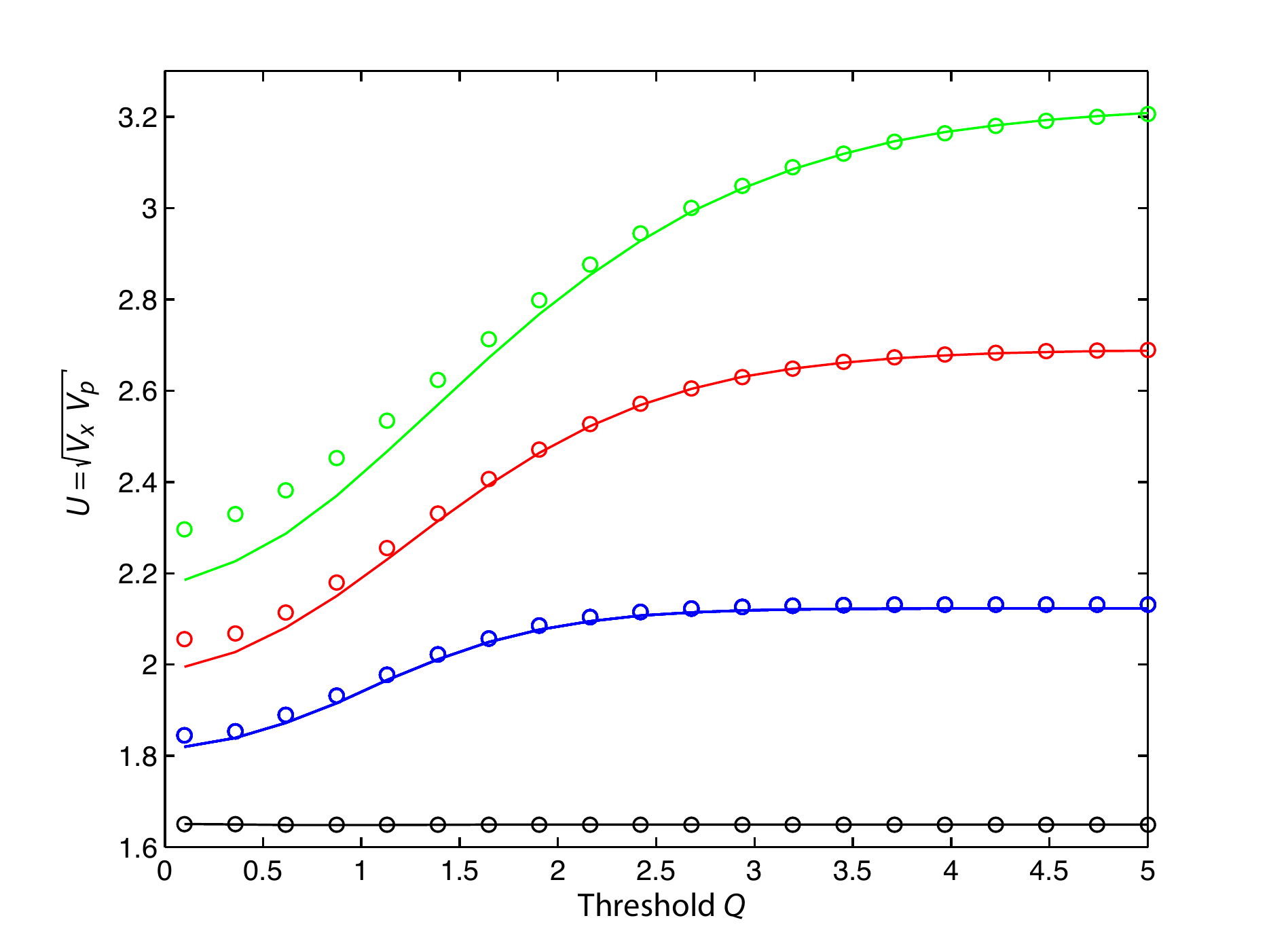}
%\centerline{\psfig{figure=fig3.eps,width=0.90\linewidth}}
\end{center}
\caption{ Uncertainty product $U=\sqrt{V_{x} V_p}$ of purified
states for different phase noise levels ($\sigma = 0.40$ (green),
$\sigma = 0.28$ (red), $\sigma = 0.17$ (blue) and $\sigma = 0$
(black)). Lines represent theoretical simulations while circles
illustrate measurements.} \label{purity}
\end{figure}

\subsection{Experimental results}
\label{sectwothree}

The variances of the squeezed and anti-squeezed quadratures of the
states employed in the experiment read $V_x=0.32$ and $V_{p}=8.5$ as
directly measured by a homodyne detector. The states generated by
the two OPAs were practically identical and the strength of the
phase noise in the two channels was always kept equal. Under these
circumstances, the variance $V_{\mathrm{in}}$ becomes equal to the
variance  of quadrature  $x_1$ (and also $x_2$) at the output of
BS$_{\mathrm{PUR}}$ and can be directly measured with HD\,I or
HD\,II of the setup shown in Fig. 1. In Fig. \ref{conjugatesigma} we
compare the conditioning on measurements of the originally squeezed
($x_1$) and anti-squeezed $(p_1)$ quadratures for a fixed threshold
$Q$ and varying phase noise $\sigma$. Shown are both the
experimental data and the corresponding theoretical predictions. We
can clearly see that the purification/distillation enhances the
squeezing and $V_{\mathrm{out}}< V_{\mathrm{in}}$. We can say that
the squeezing has been probabilistically concentrated from two noisy
de-phased copies of the state into a single copy which thus exhibits
higher squeezing. Remarkably, the conditioning on $|p_1|<Q$ not only
enhances the squeezing of the quadrature $x_2$ but for sufficiently
weak phase noise it even leads to higher reduction of fluctuations
of $x_2$ than conditioning on $|x_1|<Q$.  This is somewhat
surprising, because naively one could expect that conditioning on
$|p_1|<Q$ would rather reduce the fluctuations of quadrature $p_2$
and enhance fluctuations of $x_2$. From a semiclassical point of
view one could argue that small values of $p_1$ are detected by
HD\,I with highest probability when the phase shifts $\phi_1$ and
$\phi_2$ are such that the states impinging on BS$_{\mathrm{PUR}}$
are both squeezed in $p$ quadratures. However, this picture is
generally oversimplified. Some insight into why conditioning on
$|p_1|<Q$ can help can be gained from the expression for the output
variance (\ref{Voutconjugate}) which consists of two terms. The
first  term proportional to $B$ can be interpreted as corresponding
to the probing of the channel. In case of measurement of the
originally squeezed quadrature $x_1$, the factor
$\mathrm{erf}(Q/\sqrt{2A})$ is maximized for zero random phase
shifts. In this case $B$ attains its minimum value $B=V_x$, so the
conditioning acts as a filter that suppresses contributions
corresponding to large unwanted random phase shifts. However,
formula (\ref{Voutconjugate}) contains also second negative term,
proportional to $C^2Q$, that always reduces the variance
$V_{\mathrm{out}}$. This second distillation mechanism is of purely
quantum nature since it is a consequence of the quantum correlations
established by the interference of the two copies of the de-phased
state on the purifying balanced beam splitter BS$_{\mathrm{PUR}}$.
In case of measurements of squeezed quadrature both above mechanisms
contribute to the reduction of $V_{\mathrm{out}}$. In contrast, for
measurements of the anti-squeezed quadrature the first positive term
increases the $V_{\mathrm{out}}$  since in this case
$\mathrm{erf}(Q/\sqrt{2A})$ is maximized for phase shifts
$\phi_1=\phi_2=\pi/2$ when $B$ attains its maximum possible value
$V_p$. Thus in this case the variance $V_{\mathrm{out}}$ is reduced
solely due to the second negative term. Remarkably, this quantum
distillation mechanism is efficient enough to reduce the
fluctuations of $x_2$.
 So this conjugate purification/distillation is
a purely quantum interference effect. The purification actually
reduces the variances of both conjugate quadratures $x_2$ and $p_2$
as witnessed by the decrease of the uncertainty product
$U=\sqrt{V_{x} V_p}$, see Fig. \ref{purity}. The simultaneous
suppression of the noise in both conjugate quadratures is a
signature of the increase of purity of the state, which for Gaussian
states can be evaluated as $P=1/\sqrt{V_xV_p}=1/U$.

\begin{figure}[!t!]
\begin{center}
\includegraphics[width = 120 mm]{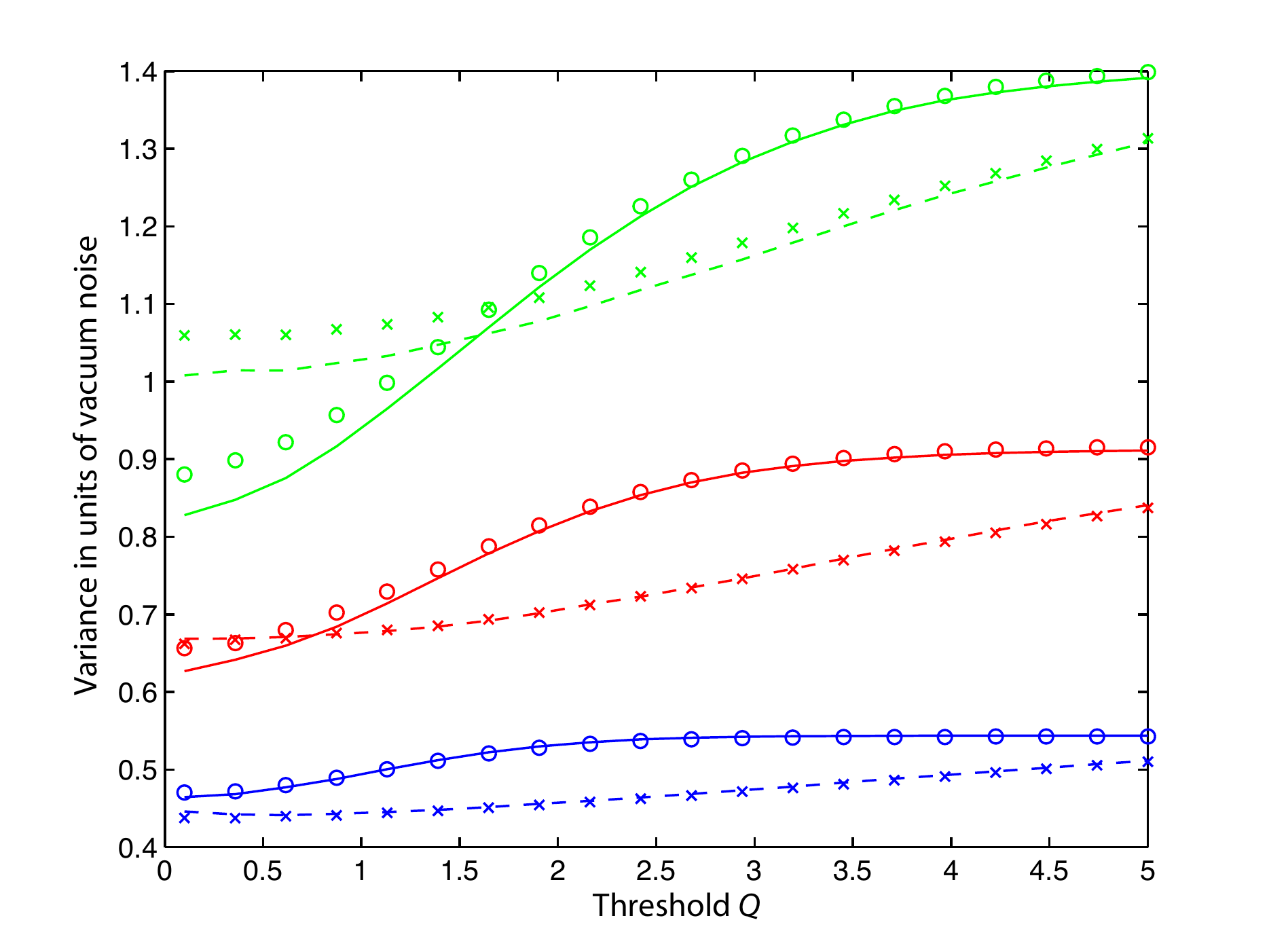}
%\centerline{\psfig{figure=fig3.eps,width=0.90\linewidth}}
\end{center}
\caption{
Quadrature variances of distilled squeezed states $V_{\mathrm{out}}$ versus trigger threshold $Q$.
$V_{\mathrm{out}}$ is plotted for three different levels of phase noise ($\sigma = 0.40$ (green),
$\sigma = 0.28$ (red) and $\sigma = 0.17$ (blue)). Results are presented for conditioning
on the (initially) squeezed quadrature $x_1$ (solid lines, circles) and for conditioning
on the antisqueezed quadrature $p_1$ (dashed lines, crosses). Again lines represent theoretical
simulations while circles and crosses illustrate measurements.
}
\label{conjugatethreshold}
\end{figure}

\begin{figure}[!t!]
\begin{center}
\includegraphics[width = 120 mm]{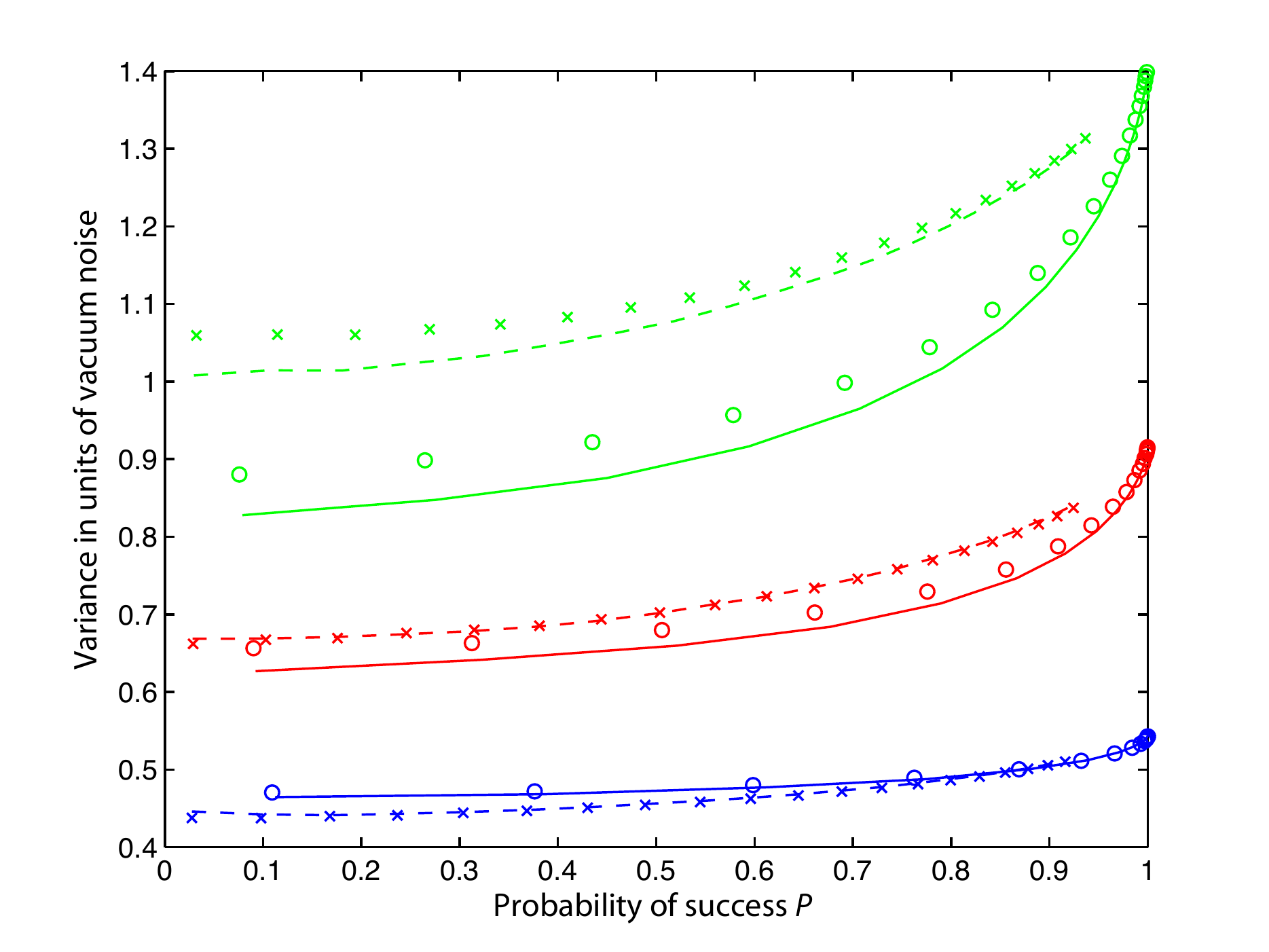}
\end{center}
%\centerline{\psfig{figure=fig4.eps,width=0.90\linewidth}}
\caption{
Illustration of the same variance data as in Fig.\,\ref{conjugatethreshold} -- this time plotted
over the fraction of distilled states. We refer to Fig.\,\ref{conjugatethreshold} for the
detailed description of parameters used. The graphs clearly show that
for weak phase noise (bottom, blue) conditioning on the antisqueezed quadrature
is more efficient than conditioning on the squeezed quadrature.
The opposite is found for strong phase fluctuations (top, green).
}
\label{conjugatesuccrate}
\end{figure}

In Fig. \ref{conjugatethreshold} we present the dependence of the
variance $V_{\mathrm{out}}$ on the trigger threshold $Q$ for three
different strengths of phase fluctuations. As expected, the output
variance decreases with decreasing trigger threshold. In agreement
with Fig. \ref{conjugatesigma} we see again that for not-too strong
fluctuations the conditioning on $p_1$ is superior to conditioning
on $x_1$ and yields lower variance $V_{\mathrm{out}}$. The
probability of success $\mathcal{P}$ monotonically increases with
$Q$ but also depends on the choice of quadrature used for
conditioning. For further comparison, we plot in Fig.
\ref{conjugatesuccrate} the trade-off between the output variance
$V_{\mathrm{out}}$ and the success rate $\mathcal{P}$. We can see
that we can achieve higher reduction of the noise at the expense of
lower success rate $\mathcal{P}$. Note also that the achieved
reduction of squeezed-quadrature variance is almost maximal already
for $\mathcal{P}$ of the order of $30\%$ and further lowering of the
success probability results only in marginal improvement of the
squeezing. This result is rather generic as confirmed by extensive
numerical simulations. We can conclude that a single iteration of
the purification/distillation procedure typically exhibits nearly
optimum performance for a rather high success probability of several
tens of percent.
 What is remarkable, for weak phase noise the conditioning on
originally anti-squeezed quadrature $p_1$ yields for a given success
rate $\mathcal{P}$ a lower variance $V_{\mathrm{out}}$ than
conditioning on the originally squeezed quadrature $x_1$. For strong
phase noise, however, it becomes preferential to condition on
measurements of $x_1$. Note that the theoretical curves obtained
from the simple theory developed in Sec. \ref{sectwotwo} are in
excellent agreement with the experimental data and the theory
faithfully describes all features of our purification/distillation
protocol.

\begin{figure}[!t!]
\begin{center}
\includegraphics[width = 120 mm]{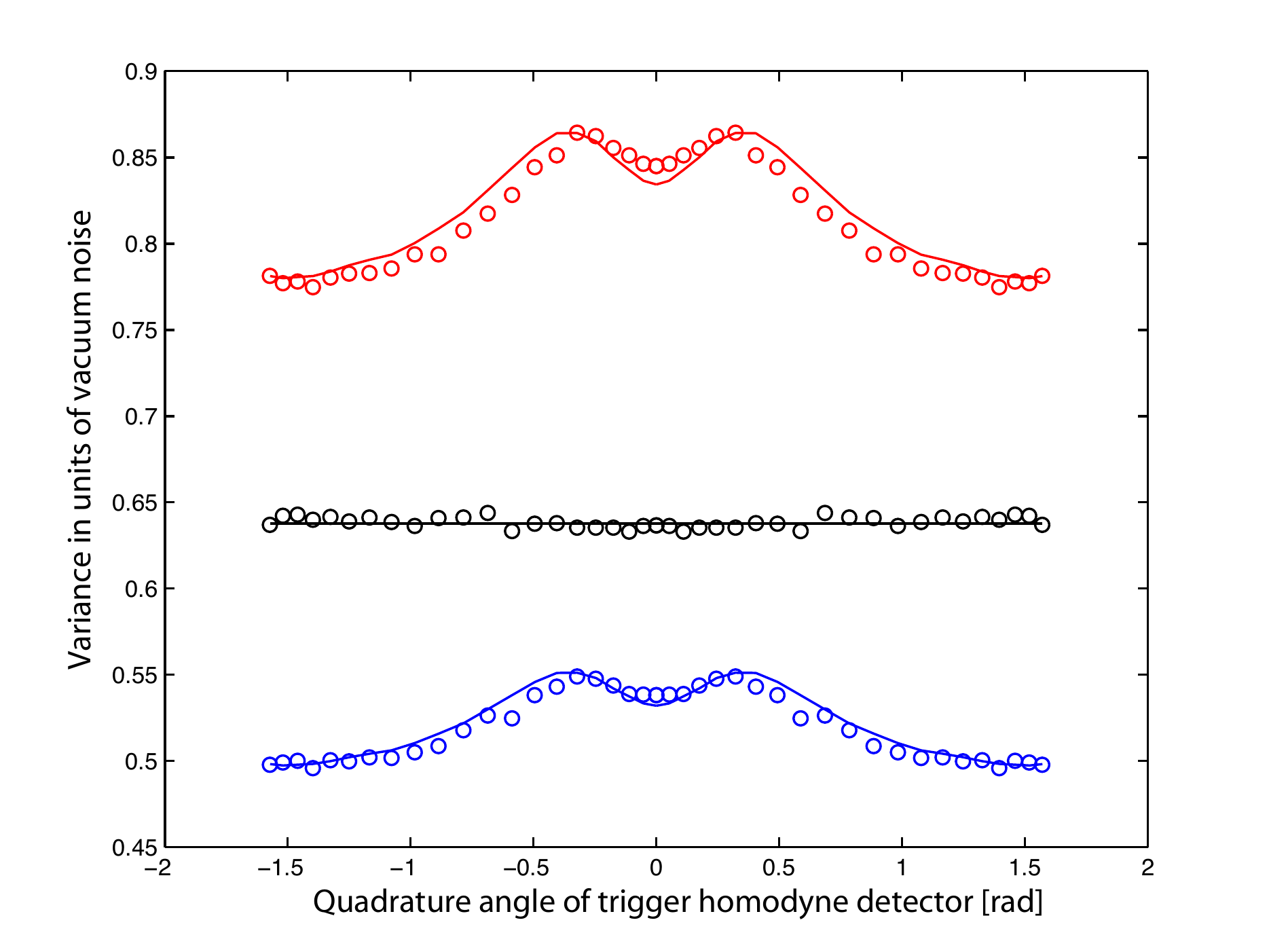}
%\centerline{\psfig{figure=fig5.eps,width=0.95\linewidth}}
\end{center}
\caption{
Experimental and theoretical characterization of our distillation protocol
for phase noise $\sigma = 0.202$ and trigger threshold $Q = 0.7$.
Shown are variances versus the conditioning quadrature angle.
The central (black) curve shows the variance of the dephased state's
amplitude quadrature $V_{\mathrm{in}}$. The lower (blue) curve shows the variance of the distilled states' amplitude quadrature $V_{\mathrm{out}}$. For this particular parameter regime conditioning on the antisqueezed quadrature works more efficiently than conditioning on the squeezed quadrature. However, the distillation protocol is successful for conditioning on \emph{any} quadrature. The top (red) curve displays the purified variance $V_{\mathrm{out}}$ after normalizing to $V_{\mathrm{in}}$ (black curve).
 }
\label{conjugatethetafig}
\end{figure}

We have also considered conditioning on measurements of arbitrary
quadrature $q_1(\theta)$ and investigated the dependence of the
performance of the purification/distillation protocol on $\theta$.
We have found that it in fact does not matter too much which
quadrature $q_1(\theta)$ is measured in the homodyne detector HD\,I
and the purification/distillation actually works well for all
$\theta$. Typical dependence of the squeezing of the purified state
on $\theta$ is depicted in Fig. \ref{conjugatethetafig}. We can see
that $V_{\mathrm{out}}$ exhibits a local minimum at $\theta=0$ but
the global minimum corresponding to the optimal
purification/distillation strategy occurs in this case at
$\theta=\pi/2$. Importantly, the quadrature fluctuations are
suppressed and the squeezing is thus enhanced for any $\theta$. This
implies that the purification/distillation works even with
phase-randomized homodyning, where the relative phase $\theta$
between balanced homodyne detector and signal is varied or randomly
fluctuates in time. Interestingly, the phase-randomized homodyning
very closely resembles the vacuum projection considered in Refs.
\cite{Browne03,Eisert04}. The effective Positive Operator Valued
Measure (POVM) element that describes this conditioning measurement
reads
\begin{equation}
\Pi_{Q}=\frac{1}{2\pi}\int_0^{2\pi} \int_{-Q}^Q |q;\theta\rangle\langle
q;\theta| \;d\theta \;dq= \sum_{n=0}^\infty P_n |n\rangle\langle n|,
\end{equation}
where $|q;\theta\rangle$ is the eigenstate of operator $q_1(\theta)$
with eigenvalue $q$ and $P_n=(\sqrt{\pi}2^n n!)^{-1}\int_{-Q}^Q
H_n^2(x)e^{-x^2}dx$, where $H_n(x)$ denotes the Hermite polynomial.
$\Pi_Q$ is diagonal in the Fock state basis because all off-diagonal
terms vanish due to averaging over random phase shift $\theta$. The
dominant part of this POVM element is the term proportional to the
projector onto vacuum state $|0\rangle$, but $\Pi_Q$ also contains
terms proportional to projectors onto higher Fock states
$|n\rangle$. This POVM can be thus considered as an approximate
noisy version of the ideal projection onto vacuum.

%         %%%%%%%%%%%%%%%%%%%%%%%%%%%%%%%%%%%%%%%%%%%%%%
%         %%                                         QUANTUM CHANNEL PROBING                                                     %%
%         %%%%%%%%%%%%%%%%%%%%%%%%%%%%%%%%%%%%%%%%%%%%%%
\section{Enhancement by Quantum Channel Probing}
\label{sectionthree}

\begin{figure}[!t!]
\begin{center}
\includegraphics[width = 120 mm]{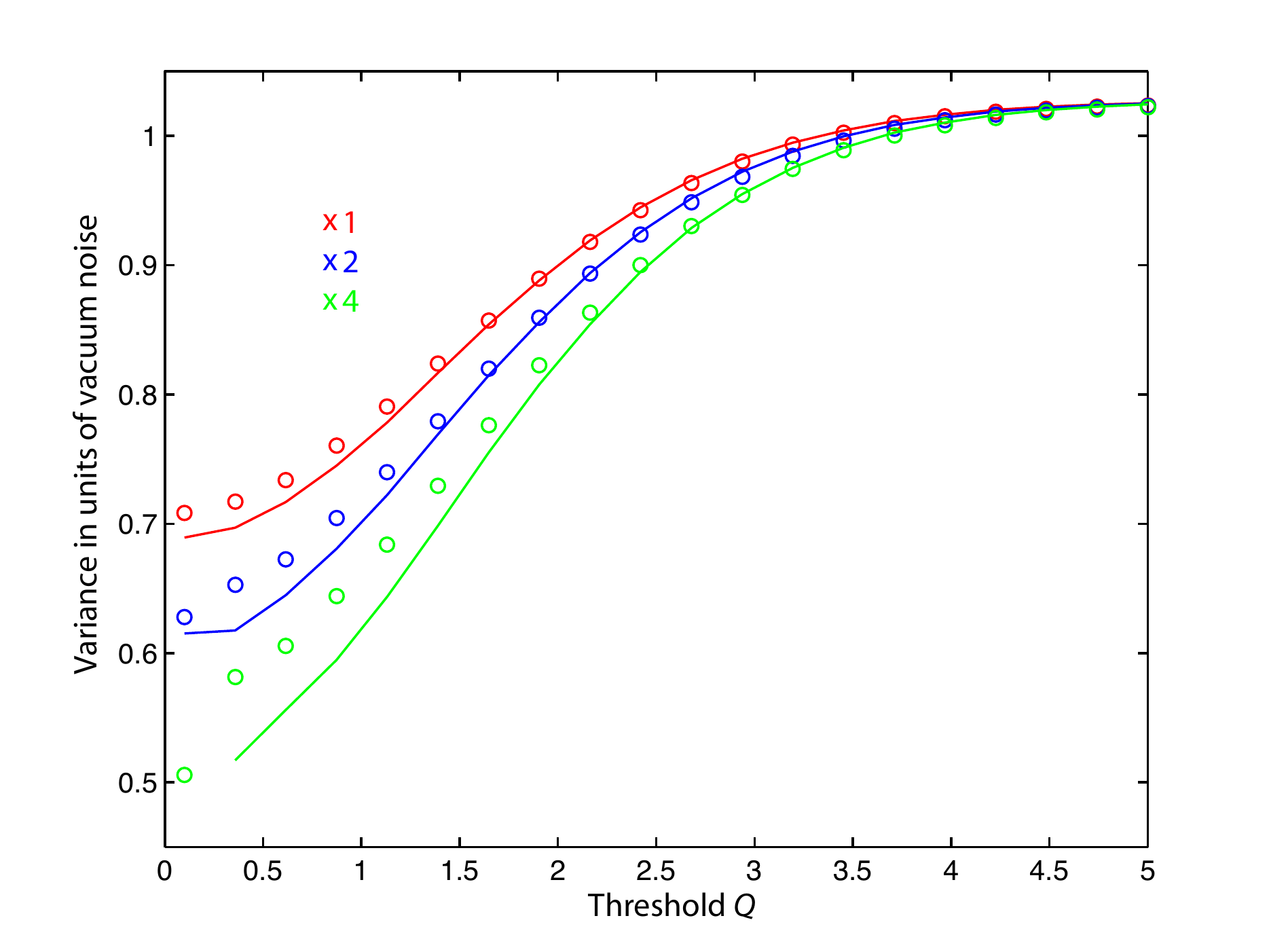}
%\centerline{\psfig{figure=fig6.eps,width=0.95\linewidth}}
\end{center}
\caption{
The effect of quantum channel probing is shown by plotting the variance of the purified
state $V_{\mathrm{out}}$ over the chosen threshold. The red curve corresponds to the previously
described method. The blue and green curves result, when conditioning is commenced
on two and four subsequent values, respectively.
 }
\label{fig6}
\end{figure}

In the previous sections we have analyzed several strategies for
conditioning in our two-copy purification and distillation protocol.
In this section we investigate a classical enhancement for our
protocol. Since a lot of copies of the decohered quantum state are
sent through the same channel they might be used as quantum probes
providing information about the channel.
%This approach cannot be used for a purification
%protocol on its own because it builds on classical information.
%However, in a realistic scenario it can be used to classically
%improve our purification protocol because phase noise (or at least
%parts of it) typically occur at frequencies well below the
%resolution bandwidth with which optical states are measured.
This approach can be used to classically improve our
purification/distillation protocol if the phase noise (or at least
parts of it) occurs at frequencies well below the resolution
bandwidth with which optical states are measured. In this case the
random phase shifts of subsequently arriving copies of squeezed
states are not completely independent. Here, we propose a protocol
which exploits these correlations between phase fluctuations of
several subsequent copies of the state which we term \emph{quantum
channel probing} (\emph{QCP}). Since QCP is an extension of the
post-processing stage of our purification/distillation protocol it
can be (and in the following is) applied to the same data as the
latter.

\begin{figure}[!t!]
\begin{center}
\includegraphics[width = 120 mm]{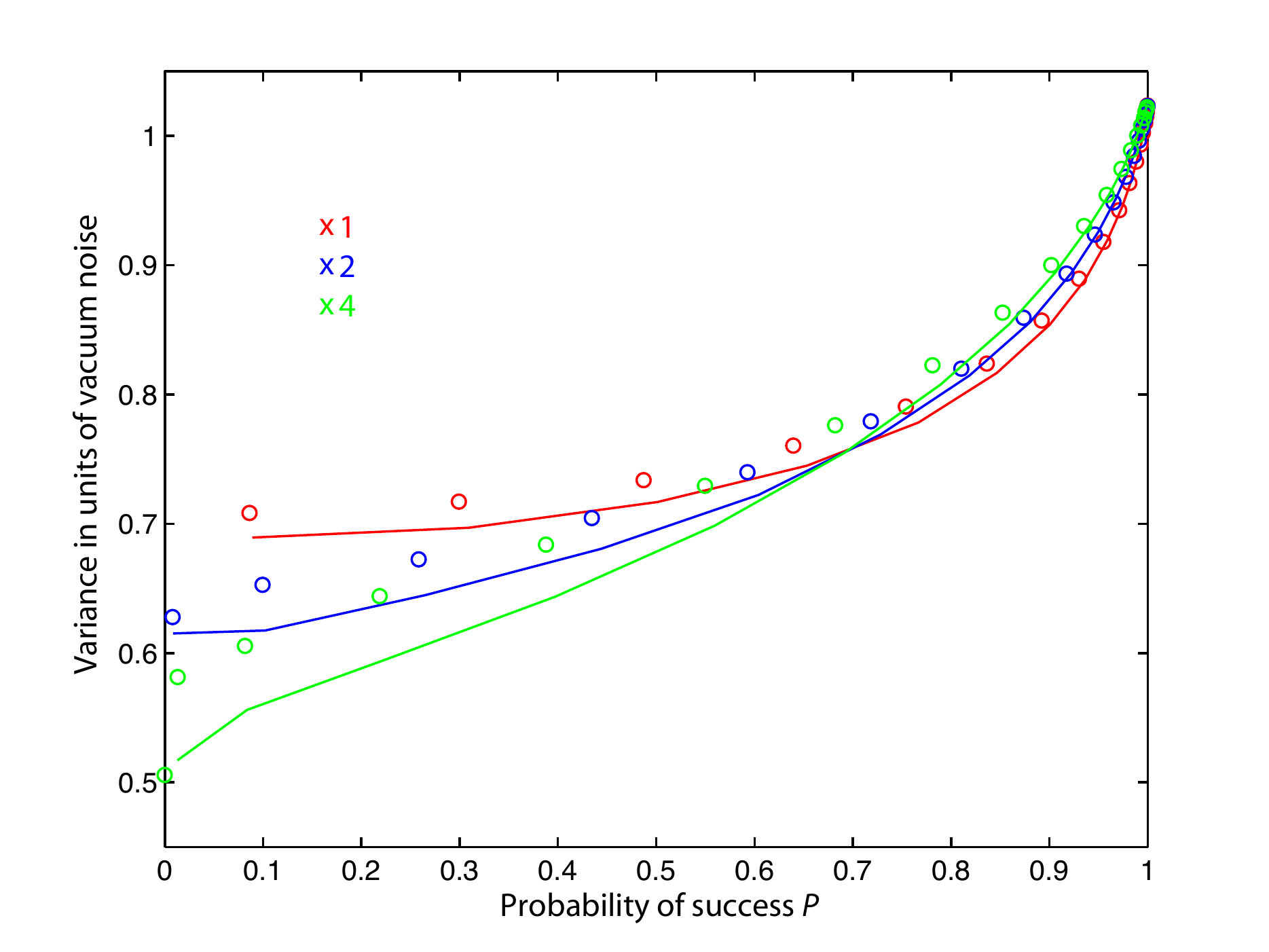}
%\centerline{\psfig{figure=fig7.eps,width=0.95\linewidth}}
\end{center}
\caption{ The same data as in figure \ref{fig6} is displayed, this
time plotted over the probability of success. It should be noted
that applying the method of \emph{quantum channel probing} does not
necessarily improve the squeezed quadrature's variance when a given
rate of survivors is to be achieved. For a probability of success
smaller than approximately $0.75$, though, quantum channel probing
yields a significant improvement.} \label{fig7}
\end{figure}

In QCP a positive trigger signal is generated at the homodyne
detector HD\,I if not only a single but a certain number
$N_{\mathrm{QCP}}$ of subsequent measurements all fulfill a certain
condition $|q_1|<Q$. Since we consider the QCP as an enhancement of
our purification/distillation protocol, the measurement setup still
corresponds to Fig.\,\ref{schemefig} and HD\,I measures the
quadrature $q_1(\theta)=x_1\cos\theta +p_1\sin\theta $. The theory
presented in Sec. \ref{sectwotwo} can be easily generalized to
describe the QCP protocol. In particular, in the limit of perfectly
correlated identical
 random phase shifts on $N_{\mathrm{QCP}}$ subsequent copies we obtain the following
expression for the output variance,
\begin{eqnarray}
V_{\mathrm{out}} &=&\frac{1}{\mathcal{P}_{\mathrm{QCP}}}\int_{\phi_1}\int_{\phi_2}
\left[B \, \mathrm{erf}\left(\frac{Q}{\sqrt{2A}}\right)-\sqrt{\frac{2}{\pi}}\frac{C^2
Q}{A^{3/2}} e^{-\frac{Q^2}{2A}} \right] \nonumber \\
& & \qquad \qquad \times \left[\mathrm{erf}\left(\frac{Q}{\sqrt{2A}}\right)\right]^{N_{\mathrm{QCP}}-1}
\Phi(\phi_1) \Phi(\phi_2) d \phi_1 d \phi_2.
\label{VoutQCP}
\end{eqnarray}
Here
\begin{equation}
\mathcal{P}_{\mathrm{QCP}}=
\int_{\phi_1}\int_{\phi_2}
\left[\mathrm{erf}\left(\frac{Q}{\sqrt{2A}}\right)\right]^{N_{\mathrm{QCP}}}
\Phi(\phi_1) \Phi(\phi_2) d \phi_1 d \phi_2.
\label{PQCP}
\end{equation}
is the probability of success of the QCP protocol. The effect of QCP
is represented by additional factors
$[\mathrm{erf}(Q/\sqrt{2A})]^{N_{\mathrm{QCP}}-1}$ in Eqs.
(\ref{VoutQCP}) and (\ref{PQCP}) compared to Eqs.
(\ref{Voutconjugate}) and (\ref{Pconjugate}).

\begin{figure}[!t!]
\begin{center}
\includegraphics[width = 120 mm]{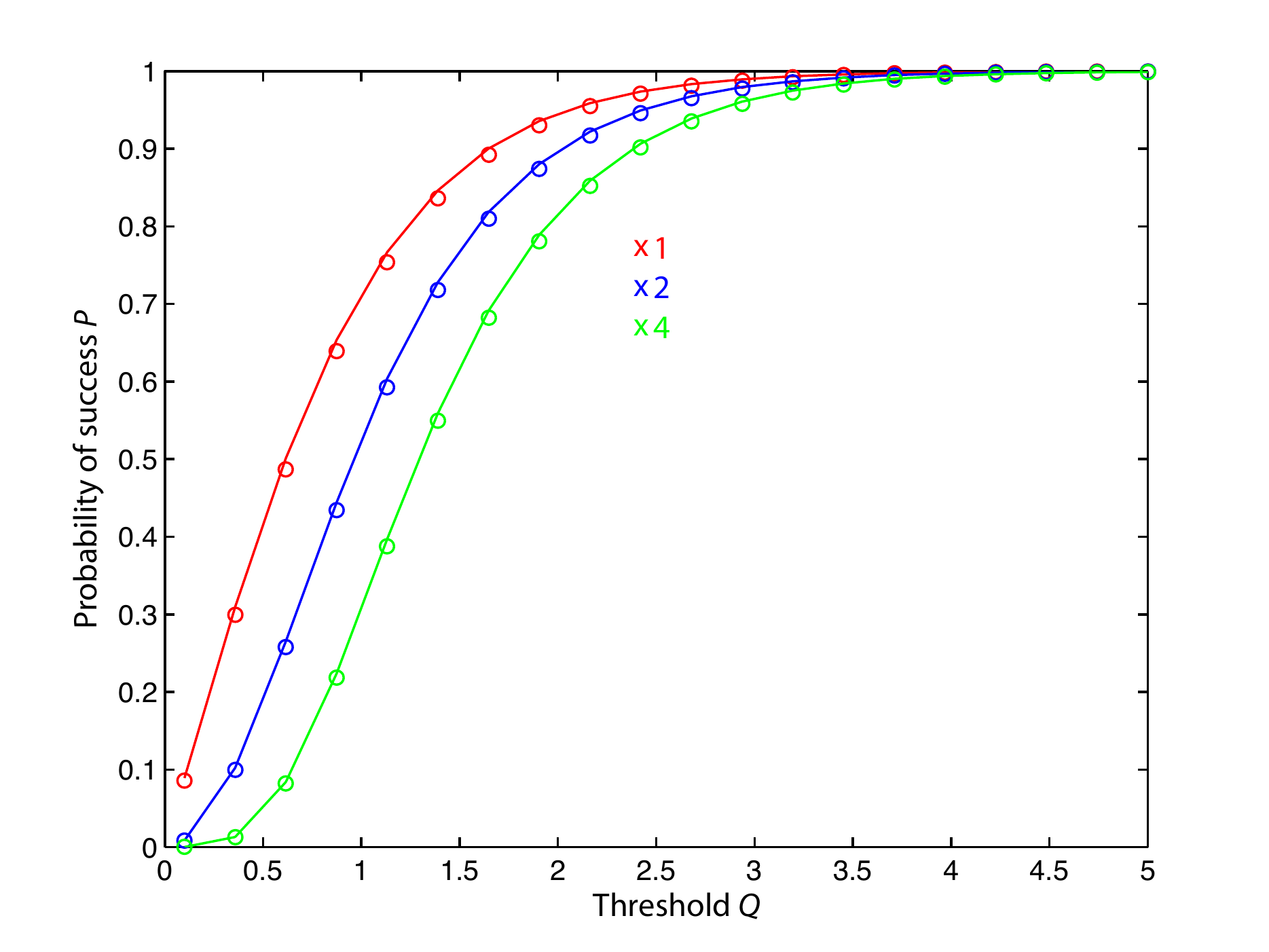}
%\centerline{\psfig{figure=fig8.eps,width=0.95\linewidth}}
\end{center}
\caption{
This figure shows the interdependence between the chosen threshold $Q$ and
the rate of survivors when the method of quantum channel probing is applied.
As before, the red, blue and green curves correspond to one, two and four
subsequent triggers (lines: theoretical predictions, circles: measured values).
 }
\label{fig8}
\end{figure}

If the observed quadrature $q_1(\theta)$ is the initially squeezed one
$q_1(0)=x_1$ this protocol probes the alignment of the squeezing
ellipses of the input states because the condition $|q_1|<Q$ for
more than one subsequent measurement is more likely met if the lean
axes of the squeezing ellipses match the quadrature of the homodyne
detector, i.e. the zero-crossing of the phase fluctuations. The
results are shown in Fig. \ref{fig6}, Fig. \ref{fig7} and Fig.
\ref{fig8}. In all cases the QCP enhanced distillation protocol was conditioned
on the initially squeezed amplitude quadrature.
Fig. \ref{fig6} shows the enhancement for $N_\mathrm{QCP}=2$ (blue)
and $N_\mathrm{QCP}=4$ (green) subsequent
triggers in comparison with $N_\mathrm{QCP}=1$, which corresponds to the original setup.
From Fig. \ref{fig8} one can see that for a
given threshold $Q$ the probability of success drops quite a bit
with increasing $N_\mathrm{QCP}$ but Fig. \ref{fig7} shows that the
improvement in the variance overcomes the decrease of success
probability.

\begin{figure}[!t!]
\begin{center}
\includegraphics[width = 120 mm]{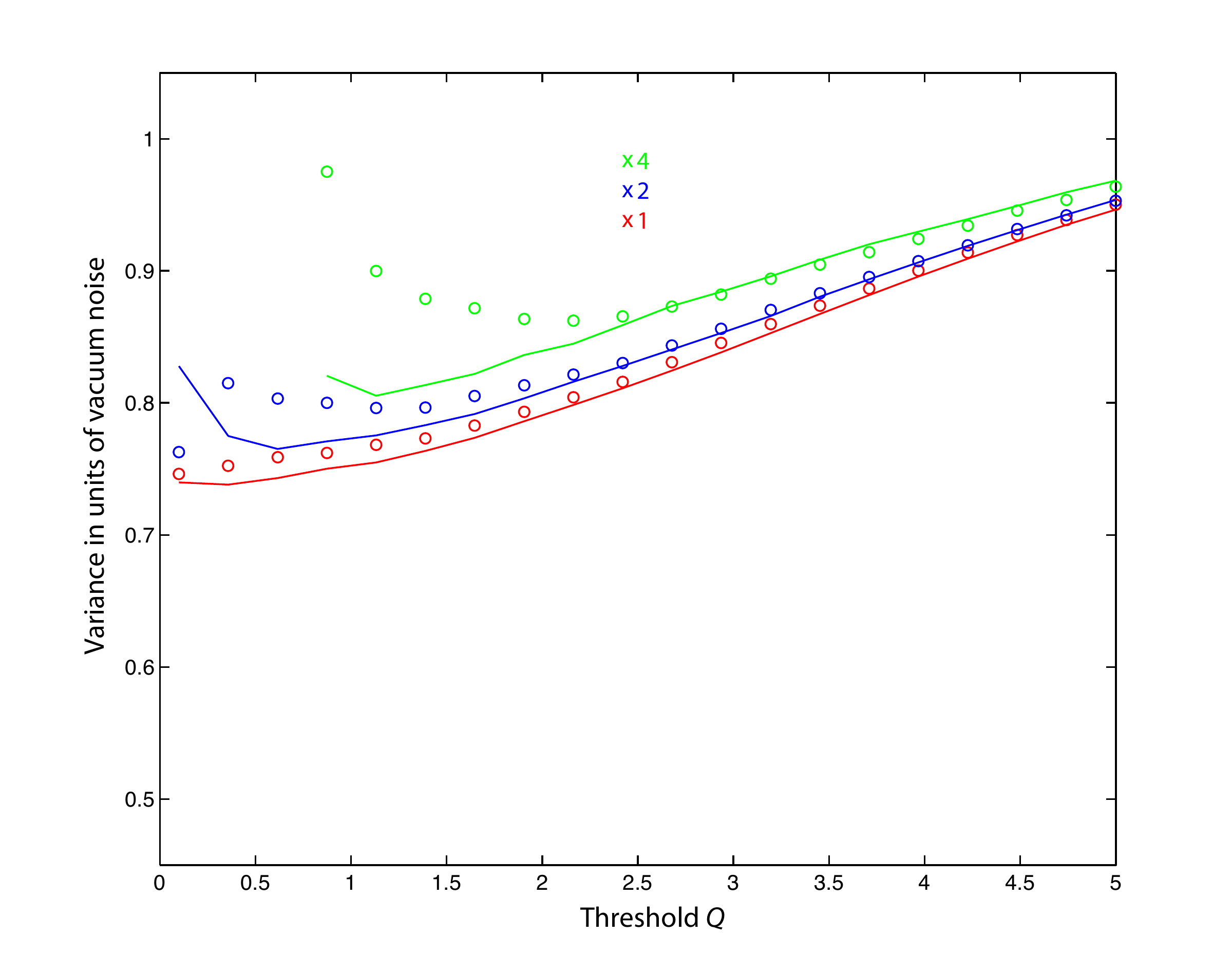}
%\centerline{\psfig{figure=fig6.eps,width=0.95\linewidth}}
\end{center}
\caption{ The QCP does not work when the trigger homodyne detector
HD\,I is tuned to measure $p_1$. Thus it is more obvious that our
protocol is more than just classical channel probing. As before, the
red, blue and green curves correspond to one, two and four
subsequent triggers (lines: theoretical predictions, circles:
measured values).
 }
\label{fig9}
\end{figure}

We emphasize that no improvement was found when the trigger was
generated from a measurement of the anti-squeezed quadrature. Fig.
\ref{fig9} shows that in this case the distillation protocol was in
fact even less efficient. This is not surprising because in this
latter case the QCP acts against the conjugate purification
mechanism represented by the negative term in Eq. (\ref{VoutQCP}),
c.f. also discussion in Sec. \ref{sectwothree}.

%         %%%%%%%%%%%%%%%%%%%%%%%%%%%%%%%%%%%%%%%%%%%%%%
%         %%                                                         CONCLUSION                                                                      %%
%         %%%%%%%%%%%%%%%%%%%%%%%%%%%%%%%%%%%%%%%%%%%%%%

\section{Conclusions}
\label{sectionfour}

We performed and analyzed an experiment that demonstrated several
strategies for conditioning the purification/distillation of
phase-diffused squeezed states. Two copies of the decohered states
were overlapped on a beam splitter and one beam splitter output was
detected with a balanced homodyne detector to provide a trigger
signal. In case of such a trigger event, states with regained
squeezing strength were found at the second beam splitter output. In
all cases we also found that the distilled states had a higher
\emph{purity}. Detection of the second beam is not required for the
protocol and the purified state is thus available for further
applications. This should be contrasted with purification of
polarization-entangled states of two photons, where the
determination of the success of the purification procedure requires
destructive detection of the purified photons \cite{singlephoton}.
Our theoretical and experimental results show that good trigger
signals could be generated from arbitrary quadratures including the
originally anti-squeezed one. This is possible due to the quantum
interference of the two copies of the phase-diffused state on the
purifying beam splitter which creates quantum correlations between
the quadratures of the two output beams. This result is also of
practical relevance. It means that the local oscillator of the
homodyne detector that produces the trigger signal does not need to
be phase locked. Hence one control loop can be saved. The
experimental result was in excellent agreement with our theoretical
predictions. Very good to excellent agreement between experiment and
theory was also found for the observed quality of our
purification/distillation protocol in dependence on the phase noise
strength, the trigger threshold value and the number of purified
states with respect to the number of input states.

We also proposed to enhance our purification/distillation protocol
by quantum channel probing. This add-on utilizes classical
correlations between decohered states transmitted through the same
channel.  With this extension the purification/distillation protocol
could further be optimized, however, the trigger quadrature had to
be the squeezed quadrature. Triggering on the anti-squeezed
quadrature failed in this case. This observation clearly draws a
line between a purification/distillation protocol that builds on
quantum correlations between two copies, and a quantum channel
probing protocol that builds on classical correlations between
states subsequently transmitted through the channel.

%         %%%%%%%%%%%%%%%%%%%%%%%%%%%%%%%%%%%%%%%%%%%%%%
%         %%                                            ACKNOWLEDGEMENTS                                                                 %%
%         %%%%%%%%%%%%%%%%%%%%%%%%%%%%%%%%%%%%%%%%%%%%%%
\ack

We acknowledge financial support from the Deutsche
Forschungsgemeinschaft (DFG), project number SCHN {757/2-1}. J.F.
acknowledges financial support from  the Ministry of Education of
the Czech Republic under the projects Centre for Modern Optics
(LC06007)  and Measurement and Information in Optics (MSM6198959213)
and  from the EU under project COVAQIAL (FP6-511004). P.M.
acknowledges financial support from the European Social Fund.
Furthermore we thank R. Filip for stimulating discussions.

%         %%%%%%%%%%%%%%%%%%%%%%%%%%%%%%%%%%%%%%%%%%%%%%
%         %%                                                         BIBLIOGRAPHY                                                                   %%
%         %%%%%%%%%%%%%%%%%%%%%%%%%%%%%%%%%%%%%%%%%%%%%%

\end{document}